Research Article

# Prevalence, Devices Used, Reasons for Use, Trust, Barriers, and Challenges in Utilizing Generative AI among Tertiary Students

John Paul P. Miranda[1*], Joseph Alexander Bansil[1], Emerson Q. Fernando[1], Almer B. Gamboa[1], Hilene E. Hernandez[1], Myka A. Cruz[1], Roque Francis B. Dianelo[1], Dina D. Gonzales[1], Elmer M. Penecilla[1]

1. **Don Honorio Ventura State University**, Pampanga, Philippines

*** Correspondence:**
John Paul P. Miranda, Don Honorio Ventura State University, jppmiranda@dhvsu.edu.ph

## ABSTRACT

This study examined generative AI usage among Philippine college students particularly on frequency, devices, reasons, knowledge, trust, perceptions, and challenges. Most students used free AI tools on smartphones due to financial constraints. They used it primarily for homework, idea generation, and research. Less than half felt confident with AI and expressed mixed feelings about its accuracy. Barriers included limited access, lack of teacher support, difficulty understanding outputs, and financial constraints. The study highlighted the need for better access, support, training, and ethical guidelines. Broader concerns included impacts on learning, academic standards, job loss, and privacy. Students viewed AI positively due to peer support. Recommendations are discussed.

# INTRODUCTION

Generative artificial intelligence (AI) creates new content like images, text, or music by learning from patterns and data (Garcia, 2024; Takale et al., 2024; E. Zhou & Lee, 2024). This technology was important because it could make new and creative things on its own, with uses in business, medicine, cloud computing, and education (Ahmad et al., 2021; Bringula, 2024; Garcia, 2024; Limna, 2022; Zheng & Wen, 2021). It could change industries by taking over creative tasks and making work more efficient and productive (Ahmad et al., 2021; Bringula, 2023; Limna, 2022; Rasul et al., 2023). However, using generative AI raised ethical questions, especially regarding who was responsible for what the AI does (Blau et al., 2024; Kim et al., 2024; Oniani et al., 2023). In education, it helps by making learning materials more accessible and tailored for each student (Javaid et al., 2023; Rasul et al., 2023; Tapalova & Zhiyenbayeva, 2022). Developing ethical guidelines, particularly in business, medicine, and education, was also important for using AI responsibly (Alam, 2023; Sand et al., 2022; Slimi & Carballido, 2023; Weidener, 2024).

# LITERATURE REVIEW

## Applications and Advancements of Generative AI

Generative AI technologies were changing many fields by offering new and creative solutions. Examples include ChatGPT, a large-scale language model developed by OpenAI, which is popular in business, medicine, education, and content creation (George et al., 2023; Khan et al., 2023; Kim et al., 2024; Matthews & Volpe, 2023). ChatGPT generated human-like text and has been used to improve medical education and clinical management (Khan et al., 2023). Another example is DALL-E, also from OpenAI, which creates images from text descriptions, showcasing its potential in creative art (Kim et al., 2024; E. Zhou & Lee, 2024). Additionally, Gemini, previously known as Bard, generates new content and is used in design, education, and healthcare (Baytak, 2024; Carlà et al., 2024; Hiwa et al., 2024; Preiksaitis & Rose, 2023). These technologies were valuable because of their ability to produce diverse outputs, streamline tasks, boost creativity, and offer personalized solutions, highlighting their importance in the AI-driven world.

## Generative AI in Educational Settings

Generative AI has had a significant impact on education by transforming traditional teaching and learning methods. It offers personalized learning experiences and improves educational resources and student outcomes (Alasadi & Baiz, 2023; Alkan, 2024; Bringula, 2024; Cooper, 2023; Maghsudi et al., 2021; Oseremi Onesi-Ozigagun et al., 2024; Tapalova & Zhiyenbayeva, 2022). AI tools change how students learn and how teachers deliver content from personalized recommendations to auto-grading essays and enhancing resources (Naseer et al., 2024; Ramani, 2022; Sadler et al., 2024; Vinutha et al., 2022). Medical education benefits from generative AI's potential, but also raises ethical and legal issues, emphasizing the necessity for AI ethics training (Jeyaraman et al., 2023; Karabacak et al., 2023; Preiksaitis & Rose, 2023). The use of AI in educational assessments could improve learning outcomes and equip students with essential 21st-century skills (González-Pérez & Ramírez-Montoya, 2022; Ng et al., 2023; Owan et al., 2023). Educators were exploring AI to provide more individualized experiences and support student learning (Elbanna & Armstrong, 2024; Foroughi et al., 2023; Tapalova & Zhiyenbayeva, 2022; Zawacki-Richter et al., 2019). Beyond traditional classrooms, AI has been used in vocational education for dynamic career planning by analyzing data to create customized content (Duan & Wu, 2024; Maghsudi et al., 2021; Tapalova & Zhiyenbayeva, 2022). AI in education was also growing, emphasizing the need for educators to trust AI-powered technology and engage in professional development (Ayanwale et al., 2024; Nazaretsky et al., 2021, 2022).

## Generative AI and the Philippine Higher Education

ChatGPT and Gemini, two leaders in generative AI technologies, are revolutionizing higher education globally and particularly in the Philippines. These tools may help students improve their writing skills (Alharbi, 2023) and boost their creativity (Garcia, 2024). The acceptance of generative AI was influenced by factors like effort expectancy, social influence, and initial trust (Gupta, 2024; Gupta & Yang, 2024; Ivanov et al., 2024; Russo, 2024; Tran et al., 2021). Students were promptly allowed to critically evaluate AI-generated text in a safe learning environment (Parker et al., 2024; Varuvel Dennison et al., 2024). Familiarity with AI concepts helps students to better understand the outputs of these tools (Alharbi, 2023; Chounta et al., 2022; Dai et al., 2023; Ng et al., 2022), and the increasing use of AI offers various strategies to enhance learning (Karabacak et al., 2023; Krajka & Olszak, 2024; Kuleto et al., 2021; Owan et al., 2023; Tapalova & Zhiyenbayeva, 2022; UNESCO, 2019). Positive student perceptions indicate growing awareness and acceptance of AI (Albayati, 2024; Labrague et al., 2023; Leoste et al., 2021; Syed & Basil A. Al-Rawi, 2023).

Understanding how tertiary students use AI was essential in the context of higher education in the Philippines. AI integration has been shown to improve student outcomes, foster collaborative learning, enhance understanding, and provide personalized instruction (Chen et al., 2020; Kaswan et al., 2024; Opesemowo & Adekomaya, 2024; Tapalova & Zhiyenbayeva, 2022). The implications of important issues attached to it like biases and misinformation (Ciampa et al., 2023; Demartini et al., 2020; Park & Kwon, 2024; Scatiggio, 2022; Shin, 2024; J. Zhou et al., 2023). However, challenges remain, such as the need for critical reflection on the pedagogical and ethical implications of AI (Bozkurt et al., 2021; Bringula, 2023, 2024; Cain et al., 2023; Crompton et al., 2024; Pedró, 2020; Sperling et al., 2024). In addition, studies stress the importance of adapting AI learning assessments and experiences (Olateju Temitope Akintayo et al., 2024; Sarshartehrani et al., 2024; Sharma et al., 2023; Tapalova & Zhiyenbayeva, 2022; Zawacki-Richter et al., 2019).

## RESEARCH OBJECTIVE

among Philippine tertiary students, focusing on prevalence, devices used, reasons for use, knowledge and trust, barriers and challenges, and perceptions of support from peers, teachers, and administrators. Understanding these factors was crucial for effective AI integration in higher education.

## METHODS

This study utilized a cross-sectional research design to determine the prevalence, devices, reasons for use, knowledge, trust, barriers, challenges, concerns, and issues, including the perceptions of peers, teachers, and school administrations in utilizing generative AI technologies among tertiary students at a state-funded university in Central Luzon, Philippines. The instrument, validated by two doctors specializing in educational technology, was subjected to a reliability test that involved 30 respondents from a similar state-funded university. Data collection was conducted from late March to early June 2024, with department chairpersons distributing the survey and encouraging student participation. An ethical statement based on the Philippine Data Privacy Law was included in the 24-item instrument. A total of 532 tertiary students responded (Table 1), with the requirement that respondents had to utilize generative AI technologies such as OpenAI's ChatGPT, Google's Gemini, Microsoft's Bing, JenniAI, Midjourney, DALL-E, Copy.AI, and Jasper in their studies for the second semester of the academic year 2023 to 2024. The average age of respondents was 20 years (SD = 2.02), with a higher percentage of female respondents (64.7%) compared to males (35%). Most respondents were in their 1st year (51.9%), followed by 3rd year (20.7%), 2nd year (17.1%), and 4th year (10.2%), with a very small percentage in their 5th year (0.2%). The most common major among respondents was Business (35.5%), followed by Education (19.5%) and Hospitality and Tourism (11.7%), with the least common areas being Arts and Sciences and Social Sciences, both at 1.7%.

**Table 1.**
*Respondents' characteristics*

| Characteristics | x̄ | % |
|---|---|---|
| Age (Mean ± SD) | 20 ± 2.02 | |
| Gender | | |
| - Male | 186 | 35 |
| - Female | 344 | 64.7 |
| - Others | 2 | 0.4 |
| Year level | | |
| - 1st year | 276 | 51.9 |
| - 2nd year | 91 | 17.1 |
| - 3rd year | 110 | 20.7 |
| - 4th year | 54 | 10.2 |
| - 5th year | 1 | 0.2 |
| Major/ area of studies | | |
| - Computer | 49 | 9.2 |
| - Education | 104 | 19.5 |
| - Business | 189 | 35.5 |
| - Engineering and design | 72 | 13.5 |
| - Arts and sciences | 9 | 1.7 |
| - Technical-vocational | 47 | 8.8 |
| - Hospitality and tourism | 62 | 11.7 |
| - Social sciences | 9 | 1.7 |

# RESULTS AND DISCUSSION

## Prevalence and Devices Used

The study showed that most respondents used free versions of generative AI technologies (93%). Only 1.7% paid for these tools, spending between PHP 500 and 2,000 (1 USD = PHP 58.77). Regarding how often they used these tools, Figure 1 indicates that 52.8% of respondents used them rarely, 24.4% used them weekly, 12.0% used them monthly, and only 10.7% used them daily. This indicates that generative AI tools are not a daily habit for most students. In the past three months, 41.4% of respondents rarely used these tools, 30.1% used them weekly, 19.4% used monthly, and 9.2% used daily. This suggests a moderate but infrequent use. Since the start of the academic year 2023-2024, rare usage is still the most common at 42.5%, with 25.2% using them monthly, 23.7% using them weekly, and only 8.6% using them daily. This shows that most respondents used generative AI technologies occasionally rather than regularly, with weekly and monthly usage being less common and daily usage being the least common. Furthermore, 85.2% of the respondents reported that they used smartphones to access these technologies. Laptops accounted for 38.9%, while desktop computers accounted for 10.9%. Only a small portion of the respondents (5.1%) reported using tablets. Interestingly, a few students (1.8%) accessed these AI technologies through less common devices, such as smartwatches and smart speakers (i.e., Google Home and Amazon Alexa).

One possible explanation for why students prefer or mostly use free versions of these tools is the general lack of affluence among students in the Philippines. Many students, particularly those in state-funded universities, have underprivileged backgrounds. The financial burden of a paid version of these tools is almost equivalent to the minimum daily wage of the workers in the country. Additionally, limited resources might contribute to the infrequent use of these tools. Another reason could be the lack of alignment between these technologies and their immediate academic needs, leading students to use them only for specific projects or assignments, rather than daily tasks. Moreover, the predominant use of smartphones for accessing these technologies reflects mobile devices' widespread availability and convenience. The higher cost of owning laptops, desktop computers, or tablets, coupled with their lack of portability, might also influence this preference. The results suggest that students have yet to fully embrace or integrate the use of generative AI technologies into their daily academic workflow, as reflected in their occasional use.

**Figure 1.**

*Generative AI technology usage patterns*

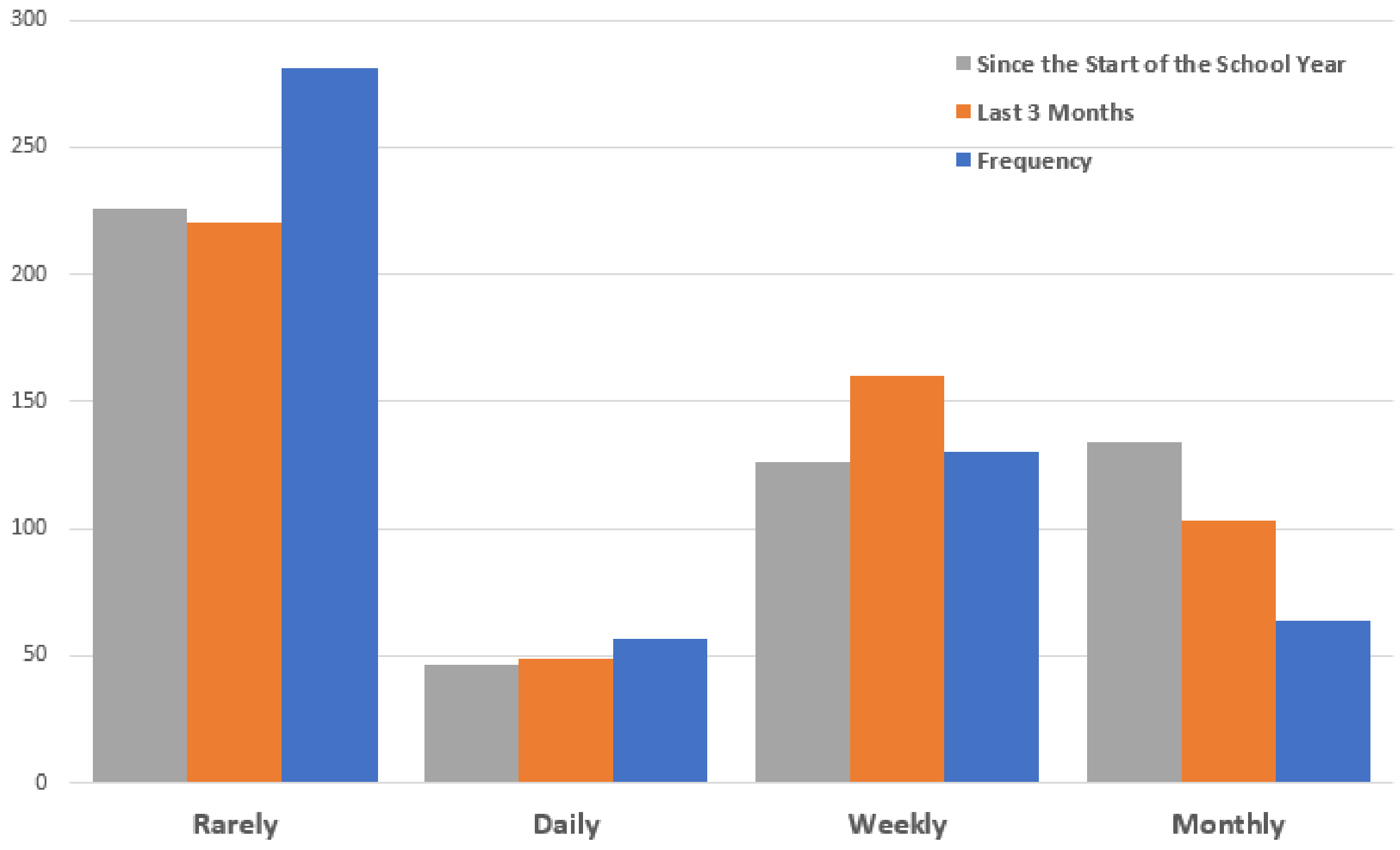

## Reasons for Use

Respondents provided a variety of motivations for using generative AI. The primary reason cited by 63.7% of the respondents was help with homework. This was followed by 46.4% using generative AI to brainstorm ideas, and 41.5% utilizing it for research purposes. Additionally, 31.4% reported using AI for language learning, whereas 28.9% used it to generate creative writing content. Furthermore, 24.8% of the respondents used AI to assist in creating presentation content, and 23.3% found it useful for study planning and organization. A smaller percentage (15 %) used AI for exam preparation and 14.7% relied on it for project management. Data analysis was another common use, reported by 13.3% of the respondents, and 12.2% used AI for programming activities. Other less common uses included helping with citations and reference styling, reviewing lessons, video game tips, creating fictional character biographies, and grammar checking. These insights highlight the diverse applications of generative AI technologies in various aspects of learning and creativity.

The results indicated that students used generative AI technology for five main reasons: academic support, idea generation, language learning, specialized academic uses, and convenience. The ability of the technology to provide answers, explanations, and summarize content saves students' time and effort in their academic work. Its ability to assist in brainstorming and improve writing styles supports its use in idea generation. Additionally, personalized language practice with real-time feedback is considered a key benefit. This indicates that the variety of specific academic assistance provided by this technology clearly explains the reasons for its use among students.

## Knowledge and Trust

The respondents showed different levels of knowledge and trust in using generative AI technologies. About 46.1% of students felt knowledgeable about using these technologies, while 41.2% felt somewhat knowledgeable, and only 2.8% felt they were not knowledgeable at all. With regard to trust, 55.1% of students had a moderate level of trust in the information from generative AI, 34% mostly trusted the information, 8.1% completely trusted it, and 2.8% did not trust it. These results suggest that while many students feel confident in using generative AI, their trust in the accuracy and reliability of information varies.

## Barriers and Challenges

The respondents reported a variety of barriers they faced when using generative AI technologies in an educational context. The primary barrier was limited access to these technologies, as cited by 44% of the respondents. Lack of support from teachers and difficulty in understanding AI outputs were the next most common barriers, both reported by 30.3% of respondents. Additionally, 21.4% mentioned a lack of peer support and resistance to adopting new technology. Insufficient training, particularly on proper prompting (18.4%), and a lack of curriculum support and integration (15.6%) were also highlighted as barriers. Accessibility issues were prevalent, with 41.4% of respondents lacking awareness of different AI tools and their potential benefits. Limited knowledge on how to use these technologies was reported by 37.2% of respondents. Language barriers, such as the need to use English for proper prompting, were cited by 27.8% of respondents. Financial constraints (23.9%) and a limited user-friendly interface (23.5%) also hindered usage. Internet access was also cited as a barrier, with 12.4% of respondents reporting limited availability of Internet at their school. Cost was another important barrier; 27.6% of respondents limited their usage to free versions due to cost, and 64.5% believed that paid services were too expensive.

Regarding challenges, the lack of a reliable Internet connection was the most important issue, reported by 52.8% of respondents. This persistent issue on the lack of reliable internet was reported by previous studies (Balahadia, 2022; Fabito et al., 2021; Javier, 2022; Narvaez et al., 2023). Software bugs and errors were noted in 39.3% of cases. A total of 23.1% of the respondents reported difficulty navigating these tools, while device compatibility issues (19.4%) and high data consumption (18.8%) were also reported as challenges. Less common challenges included a lack of proper prompt knowledge, the need for paid access, and receiving inaccurate or unreliable results. Respondents highlighted several ethical concerns. Plagiarism was the top concern, reported by 71.2% of respondents, followed by risks of misinformation (48.5%), breach of data privacy (39.3%), dependency on technology (28.6%), bias of AI outputs (25.8%), and the lack of accountability for the generated content (24.2%).

## General Issues and Concerns

The respondents reported several specific issues and concerns regarding the use of AI in education. The top concern was the difficulty in verifying the results and accuracy of the information generated by AI tools, with 46.8% of the respondents highlighting this issue. Additionally, 44.2% of respondents noted that relying on generative AI tools could diminish critical thinking skills. Many also expressed concern about becoming too dependent on AI tools, with 39.7% highlighting the potential hindrance to their learning processes. Another important issue was the difficulty in balancing the use of AI tools with traditional study methods, cited by 32.9% of respondents. Ethical dilemmas and the potential to undermine academic integrity were also raised, with 16.5% of respondents expressing this concern.

Broader concerns included the potential negative impact of generative AI on their learning and cognitive skills, which worried 52.4% of respondents. The evolving academic standards and higher expectations placed on students due to the advent of such technologies were concerns for 50.4% of respondents. The possibility of job displacement due to increased AI use was another concern mentioned by 45.3% of respondents. Privacy, security, and the lack of clear ethical guidelines for the use of generative AI were also prominent concerns, with 27.8% of respondents highlighting these issues.

### Perception of Peers, Teachers, and the School Administration

In terms of the respondents' perceptions regarding how their peers, teachers, and school administrators view the use of AI technologies in education and the support they feel from these groups. Respondents reported that their peers viewed the use of AI technologies in education positively ($\bar{x}$ = 2.87, SD = 0.654), followed by teachers ($\bar{x}$ = 2.48, SD = 0.758), and the school administration ($\bar{x}$ = 2.47, SD = 0.794). In terms of support, respondents felt positively supported by their peers ($\bar{x}$ = 2.91, SD = 0.582), teachers ($\bar{x}$ = 2.41, SD = 0.757), and the school administration ($\bar{x}$ = 2.42, SD = 0.805).

## CONCLUSION AND RECOMMENDATIONS

The heavy reliance on free versions points to the need for better accessibility and affordability of advanced features in the paid versions. The infrequent use of these tools suggests that there is potential for greater integration of generative AI in students' daily academic lives. The preference for smartphones as the primary access device highlights the importance of developing mobile-friendly interfaces for these tools. In addition to these, students reportedly use generative AI for tasks, such as homework assistance, brainstorming, research, and creative writing. Thus, these tools are essential for both academic and creative activities. A high level of self-reported knowledge indicated that students were generally comfortable using these technologies. However, varying levels of trust suggest the need to improve the perceived reliability of AI tools. Ensuring the accuracy and trustworthiness of AI-generated information is crucial for boosting student confidence and deeper integration of these technologies into academic routines.

This study also identified several key barriers and challenges faced by students when using generative AI technologies in education. The primary barrier is limited access, followed by a lack of support from teachers, and difficulty in understanding AI outputs. Other challenges include lack of peer support, insufficient training on proper prompting, and lack of curriculum integration. Additional obstacles include accessibility issues, limited knowledge of AI tools, language barriers, financial constraints, and user interface limitations. Technical challenges, such as unreliable Internet connections, software bugs, and navigation difficulties, also hinder usage. Ethical concerns, including plagiarism, misinformation, and data privacy breaches, have also been cited as challenges. Broader concerns include verifying AI-generated information, potential reduction in critical thinking skills, dependency on AI tools, and balancing AI use with traditional study methods. There are concerns about negative impacts on learning and cognitive skills, evolving academic standards, job displacement, and privacy issues. Moreover, the study revealed that students perceived slightly more support from peers than from teachers and the school administration, with peers providing the highest level of support. Furthermore, while the study provides valuable descriptive insights into these challenges. The limitations of this study include the absence of inferential statistical analysis (e.g., t-tests or chi-square tests), which potentially weakens the robustness of the results and generalizability of the insights derived from it. Future research should address these limitations to enhance the analytical rigor and provide stronger evidence for the observed trends and outcomes.